\documentclass[jgrga, draft]{agutex}

\usepackage[utf8]{inputenc}
\usepackage{epsfig}
\usepackage[T1]{fontenc}
\usepackage{graphicx}
\usepackage{amssymb}
\usepackage{amsmath}

\usepackage{relsize}
\usepackage{color}
\usepackage{bm}
\usepackage{placeins}
\usepackage{array}

\authorrunninghead{PEIXOTO ET AL.}

\titlerunninghead{SPATIOTEMPORAL CORRELATIONS OF AFTERSHOCK SEQUENCES}

\begin{document}

\title{Spatiotemporal correlations of aftershock sequences}

\author{Tiago P. Peixoto\altaffilmark{1}}
\altaffiltext{1}{Institut f\"{u}r Festk\"{o}rperphysik, TU Darmstadt, Hochschulstrasse 6,
  64289 Darmstadt, Germany}

\author{Katharina Doblhoff-Dier\altaffilmark{2}, J\"{o}rn Davidsen\altaffilmark{2}} 
\altaffiltext{2}{Department of Physics and
Astronomy, University of Calgary, 2500 University Drive NW, Calgary, Alberta,
Canada AB T2N 1N4}

\begin{abstract}
Aftershock sequences are of particular interest in seismic research since they
may condition seismic activity in a given region over long time spans. While
they are typically identified with periods of enhanced seismic activity after a
large earthquake as characterized by the Omori law, our knowledge of the
spatiotemporal correlations between events in an aftershock sequence is
limited. Here, we study the spatiotemporal correlations of two aftershock
sequences form California (Parkfield and Hector Mine) using the recently
introduced concept of ``recurrent'' events. We find that both sequences have
very similar properties and that most of them are captured by the space-time
epidemic-type aftershock sequence (ETAS) model if one takes into account catalog
incompleteness. However, the stochastic model does not capture the
spatiotemporal correlations leading to the observed structure of seismicity 
on small spatial scales.
\end{abstract}

\begin{article}

\section{Introduction}

One of the grand challenges for seismology is to establish the relationship
between stress and strain in the lithosphere~\citep{forsyth09}. While motions of
tectonic plates and surface deformations can be measured precisely with
satellite imaging and networks of Global Positioning System receivers,
strainmeters, seismometers and tiltmeters, the causative stress can only be
inferred. However, the temporally and spatially dependent rheology which
describes the linkage between the forces (stresses) and the resulting
deformations (strains) is generally not well understood~\citep{kanamori04}. This
makes it currently impossible to conclusively answer how some earthquakes
trigger other earthquakes thousands of kilometers away, for example, and to
predict earthquakes reliably.

An alternative approach to gain insight into the underlying dynamics of
earthquakes is to study the spatiotemporal patterns of seismicity where each
earthquakes is treated as a point event in space and time with a given
magnitude. Such an approach may shed light on the fundamental physics since
these patterns are emergent processes of the underlying many-body nonlinear
system. Indeed, it has been proved successful in many cases and led to the
discovery of new key features of
seismicity~\citep{rundle03,corral04,davidsen04,shcherbakov04,davidsen05m,baiesi05,felzer06,hainzl06,corral06,marsan08,zaliapin08}. For
example, in~\cite{davidsen05pm,davidsen06pm} the spatiotemporal clustering of
earthquakes in Southern California was found to show non-trivial features which
led to a new and independent estimate of the rupture length and its scaling with
magnitude. In particular, the results provide further evidence for a shadowing
effect associated with smaller
earthquakes~\citep{rubin02,fischer05,hainzl08}. The key to these findings was a
unique approach to quantify non-trivial spatiotemporal clustering based on the
view that any suitable definition of clustering should be purely contextual and
depend only on the actual history of events without any further
assumptions~\citep{davidsen06pm}. This approach utilizes the notion of
space-time records to define ``recurrences'' and maps seismicity onto a graph or
network, thus, allowing the characterization of spatiotemporal clustering by
means of tools from complex network theory~\citep{albert02,newman03}.

To elucidate the origin of the observed non-trivial clustering further, we study
here the spatiotemporal correlations of aftershock sequences which follow large
shallow earthquakes. Aftershocks are the most obvious example of earthquakes
that are triggered in part by preceding events as follows from the observed
increased seismic activity captured by the Omori law~\citep{utsu95}. Thus, their
specific clustering in space and time should provide information on the
underlying dynamics and triggering mechanisms --- see, for example,
\citep{felzer06,gomberg08}. Moreover, aftershocks are important from a
conceptual point of view since the current main paradigm in statistical
seismology classifies earthquakes as triggered events like aftershocks and
``background'' events which are hypothesized to be induced by other means. It is
important to realize, however, that the notion of background events --- and
aftershocks, for that matter --- is not well-defined as no clear physical
difference between such events has been established. In particular, background
events could be artifacts to a large extent since small earthquakes can trigger
larger events~\citep{helmstetter06,marsan08} and many small earthquakes are
typically not detected~\citep{sornette05w,sornette05w_a}.

For the aftershock sequences associated with the Parkfield earthquake (September
28, 2004) and the Hector Mine earthquake (October 16, 1999), we find that both
sequences show very similar spatiotemporal correlations as quantified by the
method of \cite{davidsen05pm,davidsen06pm}. While most features are also similar
to what has been observed previously for a 20-year catalog from Southern
California~\citep{davidsen06pm}, there is no indication of a shadowing
effect. Moreover, we find that most of the observed features can be captured by
a space-time version of the epidemic-type aftershock sequence (ETAS)
model~\citep{helmstetter02,ogata06} if catalog incompleteness~\citep{lennartz08}
is taken into account. This suggests that the spatiotemporal correlations are to
a large extent a consequence of a few established laws of seismicity \emph{and}
missing data. This was further confirmed by a comparison with a simple
non-homogeneous Poisson (NHP) model following the Omori law, but without any
spatiotemporal correlations between events. Yet, we find that this stochastic
model has several shortcomings with respect to the spatial distribution of
seismicity.

\section{Aftershock sequences and the ETAS model}

Although there is considerable statistical variability associated with
aftershocks, their behavior appears to satisfy several scaling laws to a
reasonably good approximation. Among them are the Gutenberg-Richter scaling
relation for the frequency-magnitude distribution~\citep{gutenberg} which is
certainly satisfied on long time scales~\citep{shcherbakov06}, B{\aa}th's law for
the difference between the magnitudes of a mainshock and its largest
aftershock~\citep{bath65} as well as the modified Omori law for the temporal
decay of aftershock rates~\citep{utsu95,shcherbakov04,hainzl08,narteau09}. More
recently, another scaling law characterizing the epicenter distribution of
aftershocks has been found~\citep{felzer06}.

In an attempt to establish a statistical null model of aftershocks which would
incorporate some of these scaling laws, the epidemic-type aftershock sequence
(ETAS) model was
introduced~\citep{kagan_statistical_1987,ogata_statistical_1988,helmstetter_diffusion_2002}. This
model describes a stochastic branching process in which any earthquake may
trigger other earthquakes, which in turn may trigger more, and so on. In
particular, the location and the time of occurrence of each ``daughter'' event
is strongly correlated to its ``mother'' event: The occurrence rate of daughter
events at time $t$ and position $\vec{r}$ triggered by a mother event of
magnitude $m_i$, at time $t_i$ and position $\vec{r}_i$, is defined as
\begin{equation}
  \phi_{m_i}(t-t_i, \vec{r}-\vec{r}_i) = \rho(m_i)\Psi(t-t_i)\Phi_{m_i}(\vec{r}-\vec{r}_i),
\end{equation}
where $\rho(m_i)$ is the average number of aftershocks directly triggered by an
event of magnitude $m_i$, $\Psi(t-t_i)$ is the normalized temporal distribution
of directly triggered events and $\Phi_{m_i}(\vec{r}-\vec{r}_i)$ is the
normalized spatial distribution of directly triggered events in two
dimensions. $\rho(m_i)$ is assumed to follow the productivity law
\begin{equation}
  \rho(m_i) = K 10^{\alpha(m_i-m_0)},
\end{equation}
where $m_i > m_0$ and $K$, $\alpha$ are constants. Note that the productivity
law is zero below $m_0$ implying that events smaller than $m_0$ do not trigger
other earthquakes. This condition is necessary to ensure a finite total
occurrence rate --- see \citep{sornette05w} for a discussion of its physical
relevance. $\Psi(t)$ is assumed to be determined by the modified Omori law which
can be written as
\begin{equation}\label{eq:omori}
  \Psi(t) = \frac{\theta c^\theta}{(t+c)^{1+\theta}}H(t),
\end{equation}
where $c$, $\theta > 0$ and $H(t)$ is the Heaviside step function. Note that the
exponent $1+\theta$, which describes the time distribution of the \emph{direct} 
aftershocks, is typically larger than the observed Omori exponent, which 
characterizes the whole cascade of directly and indirectly triggered 
aftershocks~\citep{helmstetter_diffusion_2002,marsan08}. Finally,
$\Phi_{m}(\vec{r})$ is assumed to follow the recently established epicenter
distribution of aftershocks~\citep{helmstetter_diffusion_2002,felzer06} which can
be expressed as
\begin{equation}\label{eq:phi_r}
  \Phi_m(\vec{r}) = \frac{\mu}{d(m)\left(|\vec{r}|/d(m)+1\right)^{1+\mu}},
\end{equation}
where $\mu=0.35$ and $d(m) = d_0 10^{0.45m}$ with $d_0 = 15$m is the rupture
length of the triggering event of magnitude
$m$~\citep{wells94,kagan02,davidsen06pm}. Note that this expression is already
problematic since it assumes a rotationally symmetric distribution of
aftershocks, thus, neglecting the typically anisotropic fault structure on which
aftershocks occur and also neglecting the frequent observation that the mother
event is located on the margin of the area of triggered events. Since the epicenters
of the aftershock sequences considered here are approximately located on a single 
fault, we focus on the one-dimensional case and discuss the two-dimensional case
as necessary.

The magnitude $m_i$ of each event is independently sampled from the normalized
Gutenberg-Richter distribution,
\begin{equation}
  P(m) = b\ln(10)10^{-b(m-m_0)},
\end{equation}
where the constant $b$ is typically close to 1 and $m_0$ is again the lower
threshold, below which no aftershocks are initiated.

An important quantity in the ETAS model is the average number of $n$ of daughter
events per earthquake, averaged over all magnitudes, which is given by
\begin{align}
  n &= \int d\vec{r} \int_{t_i}^\infty dt \int_{m_0}^\infty dm_i
       P(m_i)\phi_{m_i}(t-t_i,\vec{r}-\vec{r_i})\\
    &= \frac{Kb}{b-\alpha},
\end{align}
where the last equation assumes $\alpha < b$ which has been suggested as the
relevant regime~\citep{marsan08}.

\subsection{Data sets and parameter values}

Here, we study the aftershock sequences associated with the Parkfield earthquake
($M=6.0$, September 28, 2004) and the Hector Mine earthquake ($M=7.1$, October
16, 1999) as identified in~\cite{shcherbakov04,shcherbakov06}. In both
cases, the high-quality seismic network in the vicinity of these mainshocks
provided a particularly well-documented sequence of aftershocks. This is
especially true for the Parkfield sequence~\citep{bakun05}. For Parkfield, the
number of identified aftershock over a time period of $T=365$ days is $2056$
above magnitude $m=1.15$. For Hector mine, we have $5380$ aftershocks above
$m=2.0$. Both aftershock sequences differ significantly in their spatial extent
and also slightly in the exponents of the Omori law and the Gutenberg-Richter
law.

In principle, it seems straightforward to estimate the parameters of the ETAS
model for a given aftershock sequence. Yet, an important aspect of any
aftershock sequence is that at early times after a mainshock not all (small)
aftershocks are detected due to the large amount of seismic
noise~\citep{kagan04,kagan05,peng06}. This has led to the proposition of a
time-dependent magnitude threshold of completeness which involves further
parameters~\citep{helmstetter06}. To take this into account, we proceed as
follows: First we generate artificial aftershock sequence from the ETAS model
for a given main event with magnitude $M$ as outlined in the Appendix~\ref{sec:numerics}. Then, we
impose two constraints on the generated catalog to take into account the finite
area size covered by the empirical catalogs and to mimic missing data: i) events
that lie outside of the spatial area considered for the empirical catalogs were
discarded; ii) events below the time-varying magnitude threshold of completeness
$m_c(M,t)$ were discarded with a certain probability according to the procedure
described in \cite{lennartz_missing_2008}. To be more precise, we define
$m_C(M,t) = \max{[m_1(M,t),m_2]}$ where $m_1(M,t)=M-4.5-0.75\log_{10}(t)$
reflects the incompleteness at time $t$ after the main event of magnitude $M$,
and the time-independent sensitivity of the seismic network, different for each
empirical catalog, is captured by the constant $m_2$. Then, the probability of
observing an event of magnitude smaller than the threshold $m_c(M,t)$ is given
by $p_j(m)=10^{-\gamma_j(m_j-m)}$, where $j=1$ if $m_C(M,t)=m_1(M,t)$ and $j=2$
otherwise.

Using the parameters estimated in \cite{lennartz08} for the Parkfield and
Hector mine aftershock sequences, only $K$ and $\alpha$ of the ETAS model as
well as the fraction of background events have to be approximated. Our selection
criteria were both the total number of events in the catalogs, and the functional 
form of the total event rate (see Eq.~(\ref{eq:dressed_l})). 
The background rate was set at $1$ event/day, which gives a total event rate in
very good agreement with the empirical data (not shown). The other two
parameters were chosen, such that the average catalog size was similar to the
empirical catalogs (after some events were discarded by the procedure outlined
above), i.e., $2056$ for Parkfield and $5380$ for Hector Mine, with an accepted
deviation of at most $\pm 200$ events per catalog. In total, $100$ artificial
Parkfield and Hector Mine catalogs were generated. The parameter values used for
simulating the two different aftershock sequences are summarized in
table~\ref{tab:parameters}.
\begin{table*}
  \begin{tabular}{lccccccccccc}
    \hline\hline
    Event & $M$ & $m_0$ & $\theta$ & $c$ (days) & $b$ & $\alpha$ & K & $\gamma_1$ & $m_2$ &  $\gamma_2$ & $\lambda_b$ (days$^{-1}$)\\ \hline
    Parkfield & $6.0$ & $1.15$ & $0.09$ & $0.00395$ & $0.89$ & 0.8 & 0.2 & $0.7$  & $1.2$ & $1.5$ & 1\\
    Hector Mine & $7.1$ & $2.0$ & $0.21$ & $0.024$ & $1.01$ & 0.789 & 0.28 & $1.8$ & $1.4$ & $1.5$ & 1\\
    \hline\hline
  \end{tabular}
  \caption{Parameters used for the ETAS model.}\label{tab:parameters}
\end{table*}

\section{Network of recurrences}

We analyze the two aftershock sequences and the catalogs generated by the ETAS
model according to a recently proposed method which allows one to characterize
the spatiotemporal clustering of seismicity~\citep{davidsen05pm,davidsen06pm}. It
is based on the notion of a \emph{recurrence} in the context of spatiotemporal
point processes: An event is defined to be a recurrence of a given previous
event if it occurs closer in space to that event than all intervening events.
Recurrences are therefore \emph{record breaking events} with respect to
distance. This relationship is naturally represented by a directed network of
recurrences: Each event $a_k$ defined by its location $\vec{r}_k$ and time of
occurrence $t_k$, with $k=1,...,N$, is a vertex in the network and a directed
edge from $a_k$ to $a_{m}$ exists for $k<m$ if $a_m$ is a recurrence of $a_k$,
i.e., if the distance $|\vec{r}_m-\vec{r}_k|$ is smaller than the distance
$|\vec{r}_{k'}-\vec{r}_k|$ for all events $a_{k'}$ with $k<k'<m$. This
definition assumes that the events are ordered according to their occurrence in
time. Each recurrence, i.e. each edge on the network, is therefore characterized
by the time interval $t=t_m-t_k$ between the two connected events $k$ and $m$
and analogously by the spatial distance $l$ between the two.
Note that the mapping of the point process dynamics to the recurrence network is
entirely data-driven and does not impose any arbitrary space and time scales
other than those associated with the given event catalog.

To investigate the dependence of the network properties on the magnitude 
thresholds of events considered, we analyze networks obtained for different 
magnitude thresholds. This is crucial to identify robust features as well as 
potential scaling properties.

\section{Results}

\subsection{Topological characteristics of the network}

\begin{figure}
  \centering
  \includegraphics*[width=0.5\textwidth]{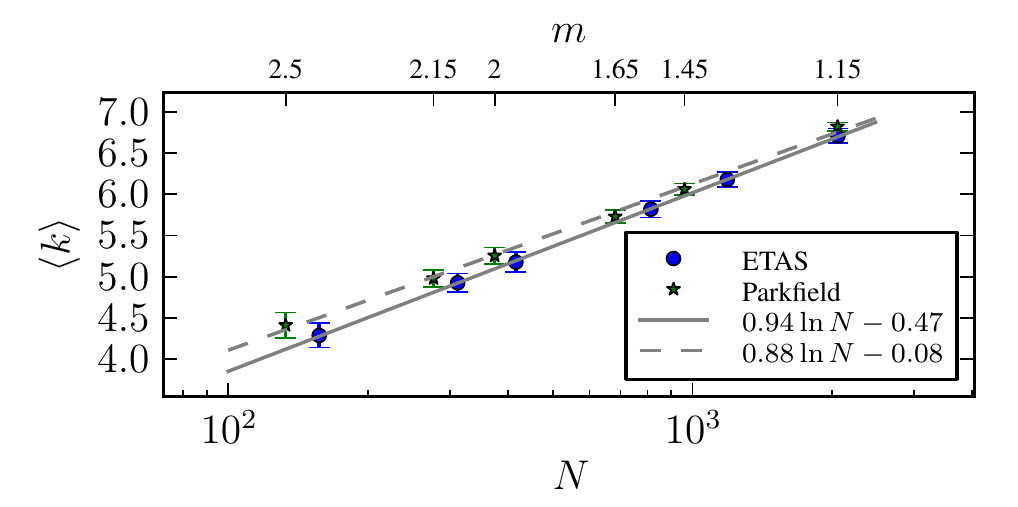}
  \includegraphics*[width=0.5\textwidth]{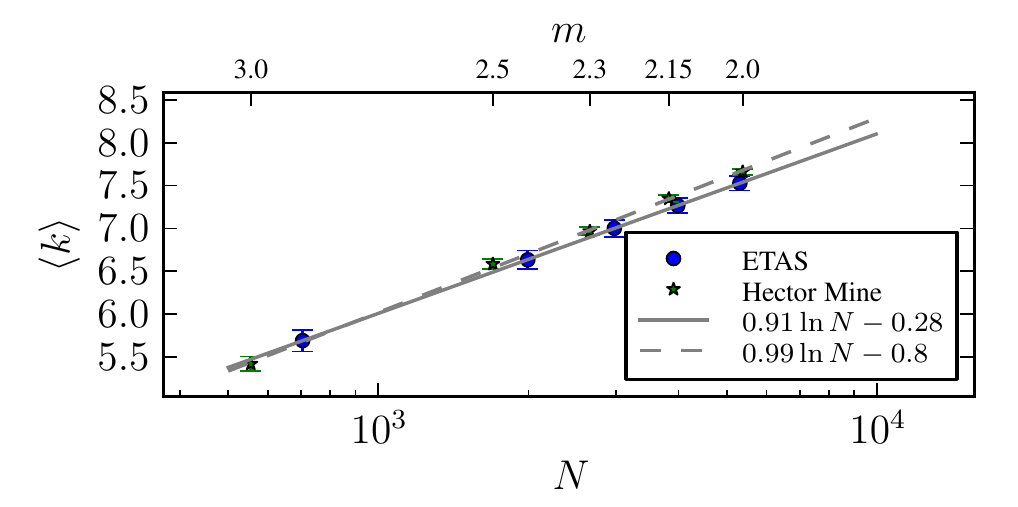}

  \caption{Average degree $\left<k\right>$ as a function of the number of events
  $N$ used to build the network. The values of $N$ were obtained by increasing
  the magnitude threshold $m$ of the event selection. The label above each plot
  indicate the corresponding $m$ values for the empirical data sets only. The
  error bars correspond to the standard deviation for the empirical datasets,
  and to the standard deviation of the average for the ETAS datasets.
  \label{fig:growth}}
\end{figure}

We turn now to the analysis of the statistical properties of the network of
recurrences which have been proved useful in the analysis of the overall seismic
activity in Southern California~\citep{davidsen05pm,davidsen06pm}. The growth of
the network can be measured by the average degree $\left<k\right>$ as a function
of the number of events $N$ in the catalog. The in(out)-degree of a node is the
number of edges pointing towards (away from) it and the in- and out-degree
averaged over the entire network are obviously the same. The number of nodes $N$
can be controlled by changing the magnitude threshold above which events are
considered for constructing the network. As can be seen in
Fig.~\ref{fig:growth}, the ETAS datasets as well as the empirical catalogs show
a logarithmic increase of $\left<k\right>$ with $N$. Both for Parkfield and
Hector Mine, the increase is compatible with the ETAS data sets if one takes
into account the statistical uncertainties. In all cases, the behaviour is
similar to what is expected if events were randomly, independently and uniformly
placed in space and time, which would imply that the growth should follow
$\left<k\right> \sim \ln N$ for large $N$~\citep{davidsen06pm}. Actually, this
result even holds if the events are not uniformly distributed in time including
the case of a simple non-homogeneous Poisson (NHP) model following the Omori law
which by construction does not have any spatiotemporal correlations between
events. The comparison with such a model is particularly helpful to identify
statistical properties of the empirical aftershock series and of series
generated by the ETAS model that arise trivially and are, thus, not due to
spatiotemporal correlations. The properties of the NHP model are analytically
derived and discussed in the Appendix~\ref{sec:nhp}.

\begin{figure}
  \centering
  \includegraphics*[width=0.5\textwidth]{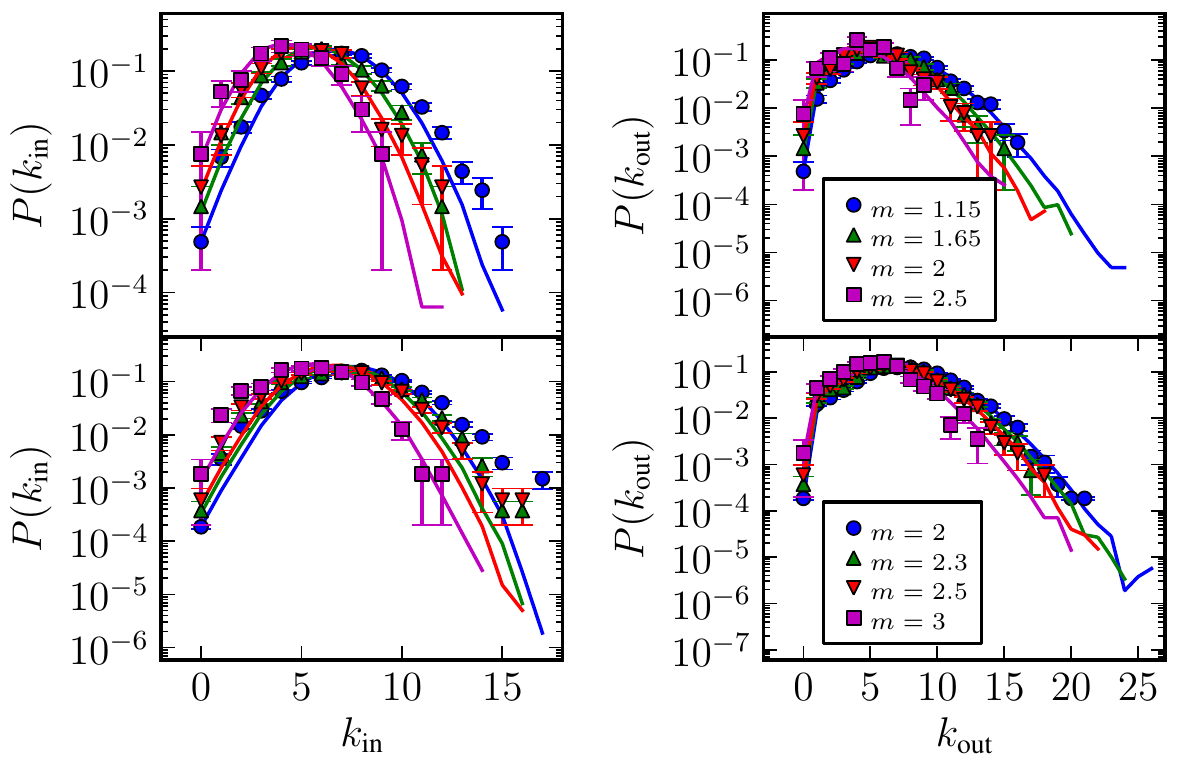}

  \caption{In- and out-degree distributions for different magnitude thresholds
  $m$, for Parkfield (above) and Hector Mine (below). Solid lines represent
  results obtained with the ETAS model, and symbols the empirical
  catalogs.\label{fig:deg-dist}}
\end{figure}

The probability distributions of in- and out-degrees can be seen in
Fig~\ref{fig:deg-dist}. The out-degree distributions for both empirical and ETAS
catalogs are in excellent agreement. While the distributions are compatible with
a Poisson distribution over the range where we have non-zero values for the
respective empirical catalog, the decay in the distribution for the respective
ETAS catalog for larger values of $k_{out}$ is slightly but significantly slower
than a Poisson distribution (not shown). The reason that we actually observe
larger values of $k_{out}$ in the ETAS catalogs is due to the much better
statistics.  The situation is different for the in-degree
distributions. Fig~\ref{fig:deg-dist} shows that there are deviations between
the distributions for the empirical catalog and for the ETAS catalog for both
Parkfield and Hector mine. The deviations are most pronounced for large $k_{in}$
and small $m$. In comparison to a Poisson distribution, the distributions are
more narrow (not shown).

For the NHP model, a Poisson distribution is expected both for the in- and
out-degree distributions in the limit of large networks. The observed deviations
imply that the degree distributions capture some of the underlying non-trivial
spatiotemporal correlations. Moreover, the differences in the in-degree
distribution between the empirical data and the ETAS model are a first
indication that the spatiotemporal correlations generated by the ETAS model
deviate from the empirical ones.

\subsection{Temporal recurrence statistics}

\begin{figure}
  \centering
  \includegraphics*[width=0.5\textwidth]{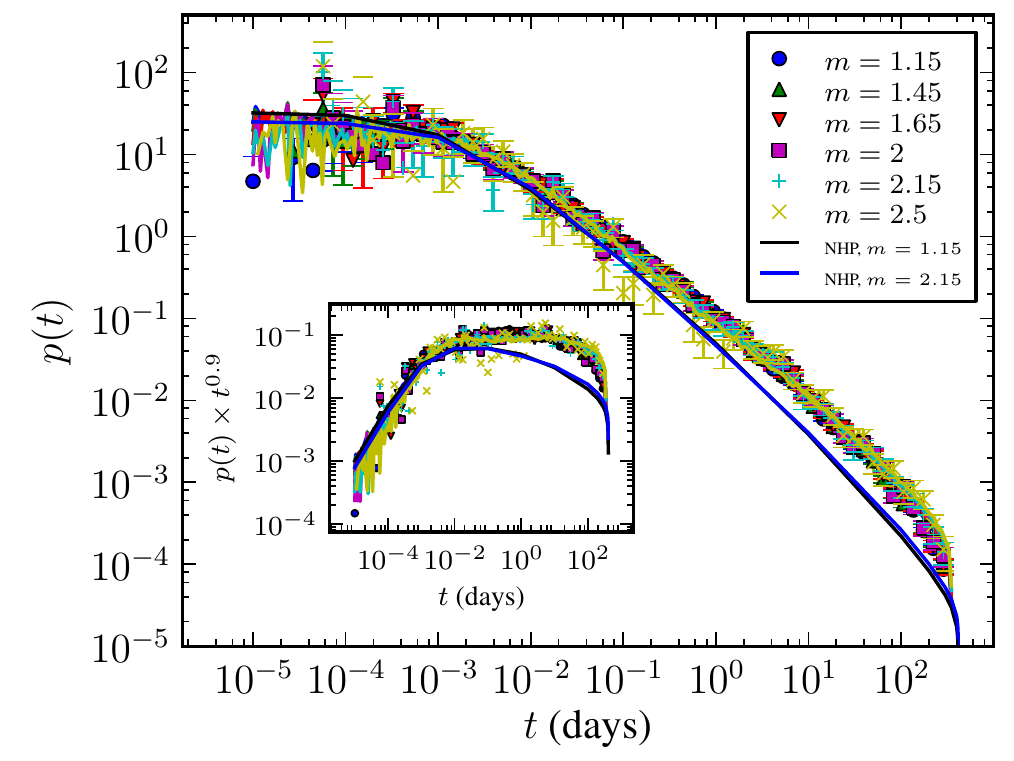}
  \includegraphics*[width=0.5\textwidth]{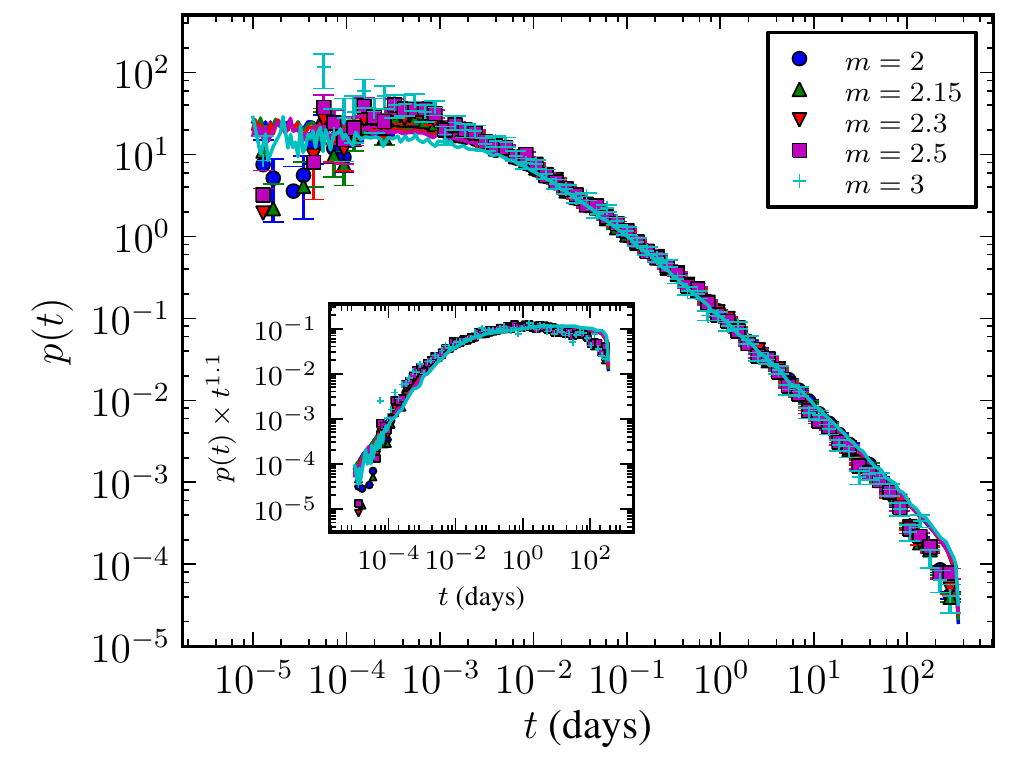}

  \caption{Probability density functions of the time intervals between an event
  and its recurrences for different threshold magnitudes $m$, for Parkfield
  (above) and Hector Mine (below). Solid lines represent results obtained for
  the ETAS model, and symbols for the empirical catalogs. The additional solid
  lines in the upper panel correspond to curves obtained with the NHP model. The
  insets show the rescaled functions as indicated by the axis labels.
  \label{fig:t-dist}}
\end{figure}

We turn now to the statistics of recurrence time intervals $t$, which are
associated with the edges of the recurrence network. The comparison between the
empirical catalogs and the ETAS model is shown in Fig~\ref{fig:t-dist}. It can
be seen that, with the parameters used for the ETAS model, and accounting
properly for the missing data, the agreement between the datasets is
excellent. Note that deviations for short time intervals are largely within the
statistical uncertainties, but a lack of resolution and missing data in the
empirical data not accounted for certainly play a role as well. This is further
supported by the fact that there are basically no significant deviations for the
Parkfield sequence which benefits from the higher quality of the seismic network
in the region. 

As the insets of Fig~\ref{fig:t-dist} indicate, after a characteristic time,
which is \emph{independent} of the magnitude threshold $m$, the probability density
functions (PDF) $p(t)$ of the time intervals decay as $t^{-0.9}$ and $t^{-1.1}$ for 
Parkfield and Hector Mine, respectively. If one does \emph{not} account for missing 
data in the ETAS models, the exponents of the power-law decay are unchanged but 
there are significant and systematic deviations at small time scales (not shown).
In particular, the characteristic time varies with $m$. This is not only further 
evidence that it is crucial to take the effect of missing data into account but it
also indicates that the characteristic time --- which is of the order of a minute 
--- might be an observational artifact.

Further support comes from the NHP model which predicts $p(t)$ as shown in the
upper panel of Fig.~\ref{fig:t-dist} for Parkfield and different magnitude
thresholds $m$ (see Appendix~\ref{sec:nhp} for details). Indeed, the apparent
lack of variation in $p(t)$ with respect to $m$ for the NHP model is due to the
dependence of the parameters in the Omori law given in Eq.~(\ref{eq:mod_omori})
on the magnitude threshold. In particular, the change in $c$ --- which is
typically related to missing data --- prevents any significant variation in the
characteristic time which would be expected otherwise (see
Eq.~(\ref{eq:fin-dom-net-pt-cases})). However, the NHP model also shows a faster
decay than what is actually observed for larger values of $t$. In fact,
Eq.~(\ref{eq:fin-dom-net-pt-cases}) predicts that we should not expect a pure
power-law for Parkfield but rather a power-law with logarithmic corrections. We
attribute this difference between the NHP model on one side and the ETAS model
and the empirical data on the other side to the presence of spatiotemporal
correlations. Clearly, this implies that at least some of the spatiotemporal
correlations present in the empirical data are correctly captured by the ETAS
model.

\subsection{Spatial recurrence statistics}

\begin{figure}
  \centering
  \includegraphics*[width=0.5\textwidth]{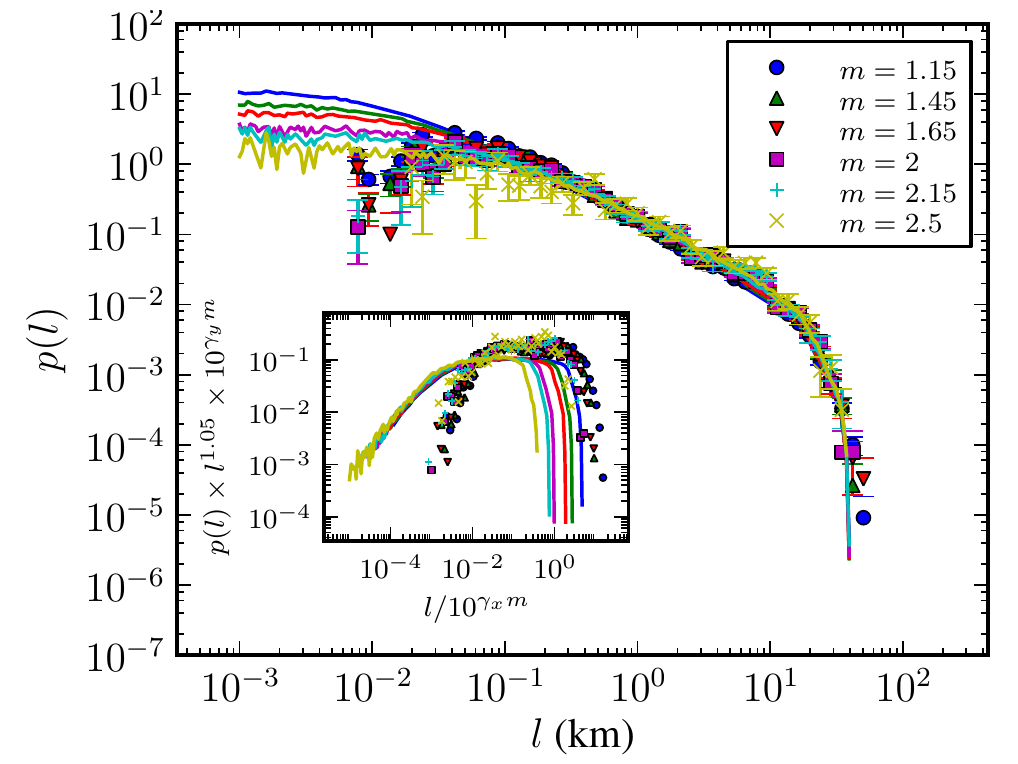}
  \includegraphics*[width=0.5\textwidth]{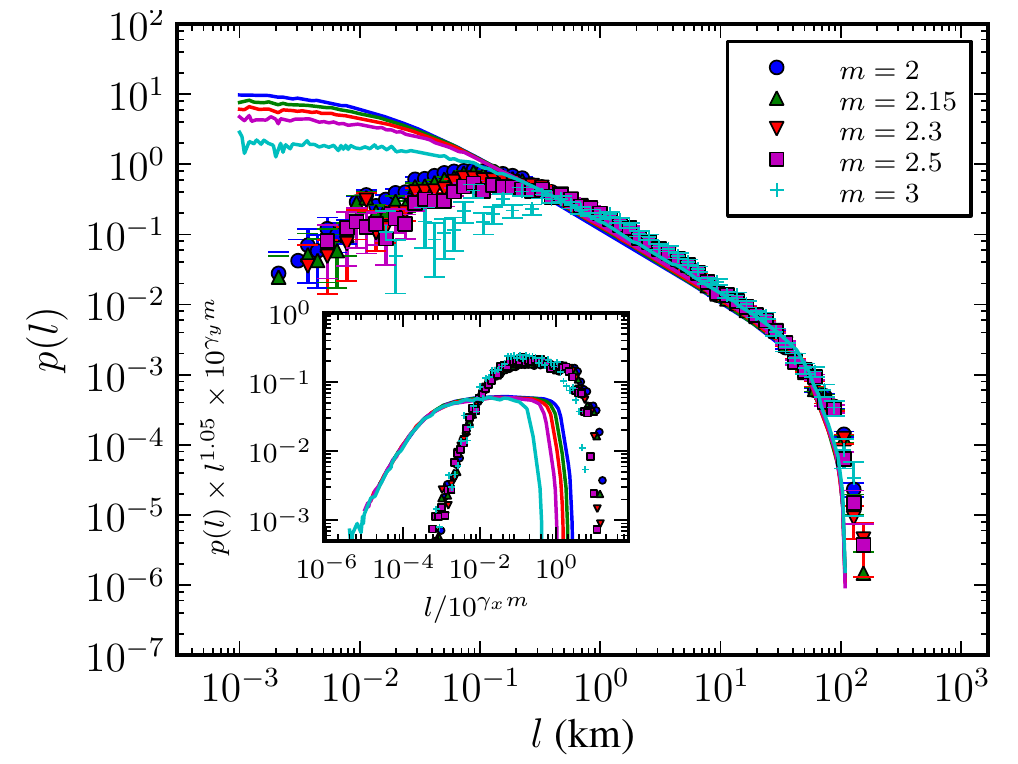}

  \caption{Recurrence distance distribution for different threshold magnitudes
  $m$, for Parkfield (above) and Hector Mine (below). Solid lines represent the
  respective ETAS model, and symbols correspond to the empirical catalogs. The
  insets show the rescaled distributions, with $\gamma_x = \gamma_y = 0.45$ for
  both empirical data sets, $\gamma_x=0.8$ for both ETAS data sets,
  $\gamma_y=-0.1$ for the ETAS model of the Parkfield sequence and
  $\gamma_y=-0.15$ for the ETAS model of the Hector Mine
  sequence\textsuperscript{\ref{fn:errors}}.\label{fig:l-dist-m}}
\end{figure}

\begin{figure}
  \centering
  \includegraphics*[width=0.5\textwidth]{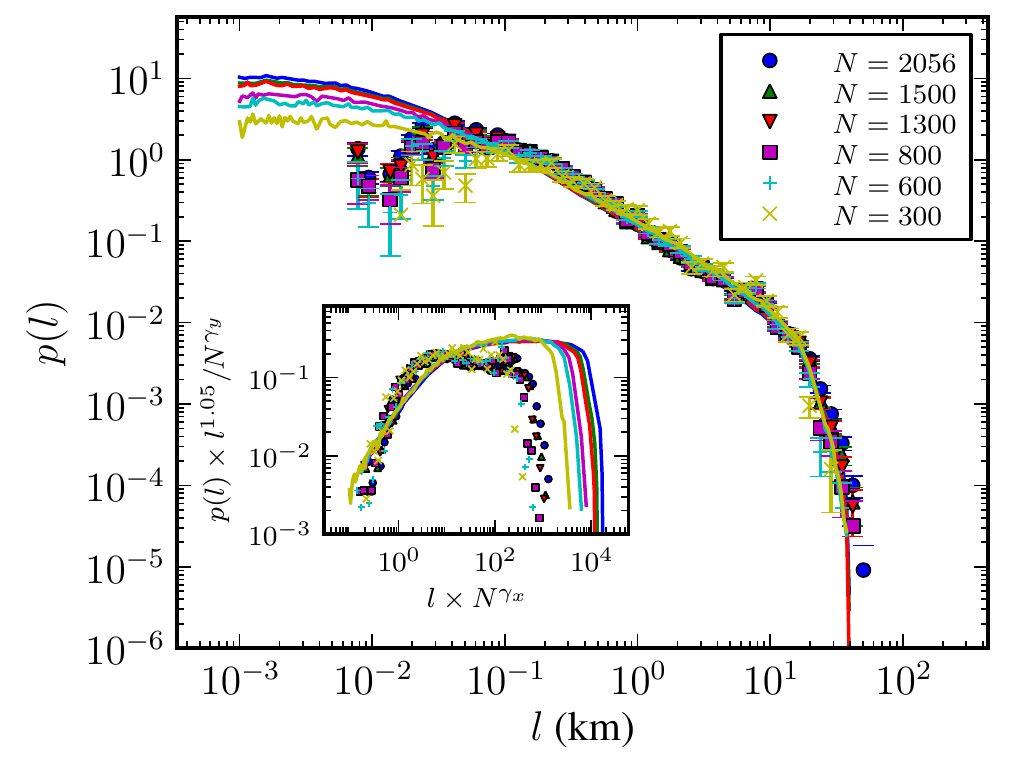}
  \includegraphics*[width=0.5\textwidth]{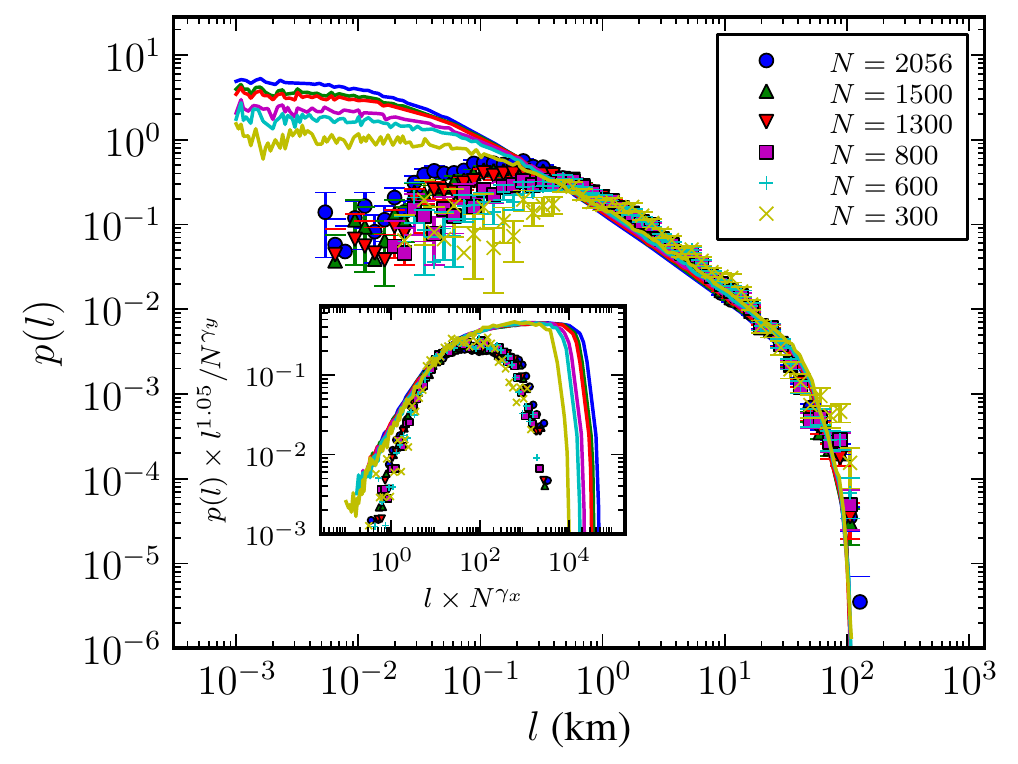}

  \caption{Recurrence distance distributions if only the first $N$ events after
  the mainshock are considered, for Parkfield with threshold magnitude $m=1.15$
  (above) and Hector Mine with $m=2$ (below). Solid lines represent the
  respective ETAS model, and symbols correspond to the empirical catalogs. The
  insets show the rescaled distributions, with $\gamma_x = \gamma_y = 0.45$ for
  both empirical data sets, $\gamma_x=0.8$ for both ETAS data sets,
  $\gamma_y=-0.1$ for the ETAS model of the Parkfield sequence and
  $\gamma_y=-0.15$ for the ETAS model of the Hector Mine
  sequence\textsuperscript{\ref{fn:errors}}. \label{fig:l-dist-n}}
\end{figure}

In addition to the time interval, there is also the spatial distance $l$
associated with each edge of the network. The PDF of these distances $p(l)$ for
both the empirical and ETAS catalogs are shown in Figs.~\ref{fig:l-dist-m}
and~\ref{fig:l-dist-n} for different magnitude thresholds $m$ and different
numbers of events $N$, respectively. Contrary to the temporal statistics, there
are significant differences between the ETAS model and the empirical data.

For the empirical data, $p(l)$ increases for small distances up to a
characteristic distance followed by a decay for larger arguments
\footnote{The relative location errors of the epicenters are typically between
10's of meters and 100's of meters for both aftershock
sequences.\label{fn:errors}}.
The decaying part itself consists of two regimes, a power-law decay for
intermediate and large distances and a much faster decay for the largest
arguments. The latter is simply a finite size effect since the maximal spatial
extent of the aftershock sequences is about $40$km for Parkfield and about
$150$km for Hector Mine. For both Parkfield and Hector mine, the power-law decay
has the same exponent of about $1.05$ as evident from the insets of
Figs.~\ref{fig:l-dist-m} and~\ref{fig:l-dist-n}. The inset of
Fig.~\ref{fig:l-dist-m} also shows that in both cases the characteristic
distance scales as $10^{0.45m}$ with magnitude threshold $m$, while it scales as
$N^{-0.45}$ with the number of considered events $N$ (inset of
Fig.~\ref{fig:l-dist-n}). Since $N \propto 10^{-m}$ according to the
Gutenberg-Richter law for $b=1$, both scaling laws are trivially related in this
case. As a comparison of the insets of Fig.~\ref{fig:l-dist-n} shows, the only
obvious difference between the aftershock sequences of Parkfield and Hector mine
are the constant prefactors in the respective scaling laws. This can be
attributed to the different geometry of the faults in the areas considered. It
is important to realize that all these findings are not significantly different
from those for the NHP model (see
Eq.~(\ref{eq:fin-dom-net-pl-cases},~\ref{eq:fin-dom-net-pl-transition}))
suggesting that they do not reflect any non-trivial spatiotemporal correlations.

For the ETAS model (see Figs.~\ref{fig:l-dist-m} and~\ref{fig:l-dist-n}), $p(l)$
shows a similar behavior for large and intermediate distances but there are huge
deviations at length scales smaller than the characteristic distance discussed
above. While there is a characteristic distance for the ETAS model, it scales as
$N^{-0.8}$ and $10^{0.8m}$, respectively
~\footnote{We note that in Figs~\ref{fig:l-dist-m} and~\ref{fig:l-dist-n}
distances smaller than $10^{-3}$km were not considered, in order to obtain a
better comparison with the empirical datasets. However, in the case of the ETAS
model, this region is statistically significant, and removing it from the
distribution destroys any chance of a symmetric collapse by a uniform scaling of
both $x$ and $y$ axes. In our case, it suffices to scale both axes
independently, where the scaling of the $x$-axis gives the characteristic length
exponent, and the $y$-axis scaling is done purely in an \emph{ad hoc} manner so
that the curves collapse.}.
This is not only different from the empirical data but there is also no obvious
relationship to the spatial propagation of activity in the ETAS model defined by
Eq.~(\ref{eq:phi_r}). Moreover, $p(l)$ does not increase for small distances but
is clearly constant. Some of these differences can be reasonably attributed to
the different dimensionality of the datasets: We have assumed for the ETAS model
that the epicenters of the events are displaced in one dimension, in order to
mimic the displacement along a single fault. However, this assumption breaks
down exactly at small distances, since on those scales the higher-dimensional
structure of the fault itself becomes important. To address this point, we have
analyzed a version of the ETAS model for Parkfield where the events are
distributed in a two-dimensional area.  While this change in dimensionality does
not affect the overall functional form of $p(l)$, the characteristic distance
now scales as $N^{-0.45}$ and $10^{0.45 m}$, respectively. Despite this
agreement with the empirical data, the prefactor in the scaling of the
characteristic distance is more than an order of magnitude larger.  It is also
important to note that $p(l)$ remains constant for small distances.  This
indicates that the spatial version of the ETAS model considered here is not able
to reliably recover the spatial distribution of aftershocks on small
scales. This is not unexpected since the isotropic spatial distribution of
triggered events used in the ETAS model does not account for the orientations of
the different faults and rupture areas. Surprisingly, however, the functional
form of $p(l)$ and the scaling behavior of the characteristic distance observed
for the empirical data can be well-reproduced by the NHP model (see
Eq.~(\ref{eq:fin-dom-net-pl-cases},~\ref{eq:fin-dom-net-pl-transition}))
\emph{despite} an isotropic spatial distribution of events in two
dimensions. This is an indication that not necessarily the isotropy of the
normalized spatial distribution of directly triggered events given in
Eq.~(\ref{eq:phi_r}) is the issue but rather those spatiotemporal correlations
that affect short spatial scales.

\section{Discussion}

It is interesting to compare the properties of the network of recurrences for
the aftershock sequences associated with the Parkfield earthquake (September 28,
2004) and the Hector Mine earthquake (October 16, 1999) with those found for a
20-year catalog from southern California~\citep{davidsen06pm}. For example, the
out-degree distributions are clearly different: While they seem to follow a
Poisson distribution for the isolated aftershock sequences, the distributions
are significantly broader for the 20-year catalog. This indicates that
spatiotemporal correlations exist between earthquakes that extend beyond what is
typically considered an aftershock sequence. Another important finding is that
the distributions of the time intervals between events and their recurrences are
basically indistinguishable. Since the analysis of the ETAS model indicates that
the absence of any dependence on the threshold magnitude for the aftershock
sequences is a consequence of missing data, one can draw an analogous conclusion
for the 20-year catalog. Finally, it is important to realize that the
distributions of the spatial distance between an earthquake and its recurrences
for the aftershock sequences do not show any indication of a non-trivial scaling
with threshold magnitude. This is in sharp contrast to the 20-year catalog,
which provided further evidence for a shadowing effect associated with smaller
earthquakes~\citep{rubin02,fischer05,hainzl08}. However, the absence of a
shadowing effect might be attributed to the fact that the main events of both
aftershock sequences studied here were of large magnitude --- for which stress
shadows are seldom observed~\citep{helmstetter05,main06} --- and, hence,
dominated the statistical behavior.

The comparison of the non-trivial statistical features of the network of
recurrences for the aftershock sequences with those for the ETAS model showed
that most of them are indeed correctly captured by the model if the effect of
catalog incompleteness is taken into account. While this is a clear indication
that some of the spatiotemporal correlations between earthquakes in aftershock
sequences can be well-understood within the framework of the stochastic model,
the significant differences in the spatial statistics of recurrences imply that
the ETAS model is inadequate to explain all spatiotemporal correlations
associated with aftershock sequences. Thus, our analysis confirms that the
statistical properties of the network of recurrences can serve as a valuable
benchmark test for models of seismicity as found in an earlier work on a
conceptual model of earthquake dynamics~\citep{peixoto07}. It remains to be seen
whether other stochastic models of seismicity pass this
test~\citep{vere-jones05,turcotte07}.

\appendix
\section*{Appendix A}
\subsection{Numerical simulation of the ETAS model}
\label{sec:numerics}

An artificial earthquake catalog can be created in an efficient manner with the
ETAS model as follows. Starting from an initial event of magnitude $M$ at the
origin, and at time $t=0$, the number of daughter events is sampled from a
Poisson distribution with average $\rho(M)$. For each daughter event $i$, its
magnitude, position, and occurrence time are independently sampled from
$P(m_i)$, $\Phi(\vec{r}_i,m_i)$ and $\Psi(t_i-t)$, respectively. The same
procedure is then repeated for each daughter event, recursively, until there are
no more events to be generated up to a predetermined maximum time. Background
events, which are not daughters of any other event, can also be added, and their
daughters can be obtained in the same manner. Thus, the total number of
iterations is proportional to the number of events in the final catalog. The
final rate of occurrence at a given time, considering all daughter events, is
then given by
\begin{equation}\label{eq:dressed_l}
  \lambda(t) =  \lambda_b + \sum_{t_i<t} K 10^{\alpha(m_i - m_0)} \frac{\theta
  c^\theta}{(t-t_i+c)^{1+\theta}},
\end{equation}
where $K=n(b-\alpha)/b$, $\lambda_b$ is the constant rate of background events, and the
sum is taken over all previous events in the catalog.

\subsection{Simple non-homogeneous Poisson model}
\label{sec:nhp}
We follow the same approach as in \cite{davidsen06pm} which allows us to define
a simple non-homogeneous Poisson (NHP) model by specifying the functions
$\mu_l(l)$ and $\mu_t(t)$. These functions quantify the spatial and temporal
distribution of events with respect to some reference event, respectively. Note
that this fact trivially excludes the existence of any spatiotemporal
correlations between events. In contrast to~\citep{davidsen06pm}, we assume here
that the rate of activity is not constant but given by the modified Omori law,
\begin{equation}
  n\left(t\right)=\frac{K}{\left(t+c\right)^{p}},
  \label{eq:mod_omori}
\end{equation} 
where we restrict ourselves to the case $p>1$ as observed for
Parkfield~\citep{shcherbakov06}. We note that $p$ is not the same as $\theta+1$
in Eq.~(\ref{eq:omori}) for the ETAS model, since the latter describes the
rate of events \emph{directly} triggered by a single event, while the former describes the
total event rate. Despite this non-homogeneous rate of activity, some of the
statistical properties of the network of recurrences remain unchanged, including
the topological structure, since they only depend on the spatial
characteristics of the point process and the temporal ordering~\citep{davidsen06pm}. 
Incorporating further the effect of finite space-time domains, we have
\begin{gather}
  \label{eq:mul-fin-dom}
  \mu_l(l)=2al\Theta\left(L-l\right),\\
  \label{eq:mut-fin-dom}
  \mu_t(t, t_0)=\frac{K}{\left(c+t+t_0\right)^p}\Theta\left(T-t-t_0\right).
\end{gather}
Here, $a$ is some constant such that $\mu_l(l)$ describes a homogeneous
distribution of events in a two-dimensional ball of radius $L$ around the
reference event. Hence, we assume translational invariance in space.  $T$
corresponds to the finite observation period after the mainshock occurring at
time $t=0$. $t_0$ is the time of occurrence of the reference event.  Using the
formalism presented in~\cite{davidsen06pm}, one can compute the average PDFs of
finding a recurrence at distance $l$ from or at time $t$ after a randomly chosen
reference event
\footnote{It is important to realize that the average has to take into account
that the rate of activity decays with time}.
This leads to 
\begin{gather}
	\label{eq:fin-dom-net-avpl}
	p_l(l) = \frac{2}{l}-\frac{2}{l^3}
		\frac{\left(p-1\right)}{a K}\frac{1}{\frac{1}{c^{p-1}}-\frac{1}{\left(c+T\right)^{p-1}}}
		\left[1-
		e^{-l^2aK\frac{1}{p-1}\left(\frac{1}{c^{p-1}}-\frac{1}{\left(c+T\right)^{p-1}}\right)}\right]
		\Theta\left(L-l\right),\\
		\label{eq:fin-dom-net-pt}
	p_t(t) =
		\frac{\left(p-1\right)}{\frac{1}{c^{p-1}}-\frac{1}{\left(c+T\right)^{p-1}}}\int_0^\infty
	 	\frac{p-1}{\left(c+t_0\right)\left(c+t+t_0\right)^p-\left(c+t_0\right)^p\left(c+t+t_0\right)}
	 	\times\\ \nonumber
		\qquad\qquad\qquad\times\left[1-
		e^{-L^2aK\frac{1}{p-1}\left(\frac{1}{\left(c+t_0\right)^{p-1}}-
		\frac{1}{\left(c+t+t_0\right)^{p-1}}\right)}\right]
		\Theta\left(t-t-t_0\right)
	 	\,\mathrm d t_0. 	 	
\end{gather}
The functional behavior of these PDFs can be broken down into several parts. For $p_l(l)$, we find
\begin{gather}
	\label{eq:fin-dom-net-pl-cases}
	p_l(l)=
	\begin{cases}
	l\frac{aK}{\left(p-1\right)}\left(\frac{1}{c^{p-1}}-\frac{1}{\left(c+T\right)^{p-1}}\right)
		& l\ll l^*\\
	\frac{2}{l}&l^*\ll l<L\\
	0 & l>L,
	\end{cases}
\end{gather}
where the characteristic distance $l^*$ scales as
\begin{gather}
	l^*\propto
		\left[aK\frac{1}{2\left(p-1\right)}\left(\frac{1}{c^{p-1}}-
		\frac{1}{\left(c+T\right)^{p-1}}\right)\right]^{-\frac{1}{2}} \propto \frac{L}{\sqrt{N/2}}. \label{eq:fin-dom-net-pl-transition}
\end{gather}
Note that the total number of events $N$ is simply given by $N = a L^2 \int^T_0 n(t) dt$. A comparison with the results derived in Ref.~\citep{davidsen06pm} for a homogeneous rate of activity shows that the qualitative features and the scaling exponents are unchanged. For $p_t(t)$, we have
\begin{gather}
	\label{eq:fin-dom-net-pt-cases}
	p_t(t) \approx
	\begin{cases}
		\frac{p-1}{\frac{1}{c^{p-1}}-\frac{1}{\left(c+T\right)^{p-1}}}
			\frac{L^2 a K}{2p-1}\left(\frac{1}{c^{2p-1}}-\frac{1}{\left(c+T\right)^{2p-1}}\right)
			&t\ll c\quad \wedge\quad t<\frac{c^p}{L^2 a K}\\
		\frac{1}{t} & t\ll c \quad\wedge\quad t>\frac{(c+T)^p}{L^2 a K}\\
		\text{see Eq.~(\ref{eq:large-values})}&t>c \quad \wedge\quad t<\frac{T+c}{2}\quad \wedge\quad t<T\\
		0&t>T.
	\end{cases}
\end{gather}
\begin{gather}
\label{eq:large-values}
p_t(t) \approx \frac{\left(p-1\right)^2}{\frac{1}{c^{p-1}}-\frac{1}{\left(c+T\right)^{p-1}}}\left\{\frac{1}{t^p}\ln\left(\frac{t}{c}\right)+\frac{1}{\left(p-1\right)^2}\left[\frac{1}{t^p}-\frac{1}{t\left(T-t+c\right)^{p-1}}+\right.\right.\\ \nonumber
		\left.\left.\frac{p-1}{pt}\left(a L^2 K t\right)^{\frac{1-p}{p}}\left[\Gamma\left(\frac{p-1}{p}, a L^2 K \frac{1}{t^{p-1}}\right)-\Gamma\left(\frac{p-1}{p}, a L^2 K \frac{t}{\left(T-t+c\right)^{p}}\right)\right]\right]\right\},
\end{gather}
where $\Gamma\left(\cdot,\cdot\right)$ denotes the incomplete Gamma function
\footnote{The above approximation for $t\gg c$ is based on splitting the
integral into two parts, $\int_0^\infty \ldots \,\mathrm dt_0=\int_0^{t-c}\ldots
\,\mathrm+\int_{c-t}^\infty \ldots \,\mathrm.$. The first term was then
approximated assuming $(c+t_0) \ll t$ and taking only terms of first order for
$(c+t_0)$ into account. The second term was approximated assuming $(c+t_0)\gg
t$, again taking only first order terms of $t$ into account.}.
Note that depending on the exact parameter values the $1/t$ regime might be present or not
\footnote{There are more cases that can be considered for $p_t(t)$, especially
one for $t>(T+c)/2$, which describes a sharp cutoff near $t$. This case,
however, only occurs over one order of magnitude and is therefore not important
for a scaling-comparison with empirical data.}.
For the aftershock sequence of Parkfield, one can also obtain $p_t(t)$ by
numerically integrating Eq.~(\ref{eq:fin-dom-net-pt}). The values of the
constants in Eqs.~(\ref{eq:mul-fin-dom},\ref{eq:mut-fin-dom}) can be derived
from the values estimated for the Omori law given in
~\cite{shcherbakov06}. Note that the values of $c$ and $K$ vary with the
magnitude threshold $m$. The variation in $c$ can arise due to missing data ---
see~\cite{shcherbakov04,shcherbakov06}, however, for a different
interpretation. For $m=2.15$, we have $c=0.016$ d and $aL^2K=21$ d$^{p-1}$. For
$m=1.15$, we have $c=0.08$ d and $aL^2K=200$ d$^{p-1}$. These curves are plotted
in Fig.~\ref{fig:t-dist}.

For $T \to \infty$ and $L \to \infty$, Eqs.~(\ref{eq:fin-dom-net-pt},\ref{eq:fin-dom-net-pt-cases}) reduce to
\begin{gather}
	p_t(t)=
		\left(p-1\right) c^{p-1}
		\int_0^\infty
		\frac{p-1}{\left(c+t_0\right)\left(c+t+t_0\right)^p-\left(c+t_0\right)^p\left(c+t+t_0\right)}
		\,\mathrm d t_0.\label{eq:pt-inf-dom-net}
\end{gather}
\begin{gather}
	\label{eq:inf-dom-net-pt-cases}
	p_t(t) \approx
	\begin{cases}
		\frac{1}{t}&t\ll t^*\\
		\frac{\left(p-1\right)^2 c^{p-1}}{t^{p}}
		\left[\ln\left(\frac{t}{c}\right)+\frac{1}{\left(p-1\right)^2}\right] &t\gg t^*,	
	\end{cases}
\end{gather}
where $t^*\propto c$.

\begin{acknowledgments}
We thank M. Werner and R. Shcherbakov for helpful discussions. This work was
supported by the DFG under contract No. Dr300/5-1, and by NSERC.
\end{acknowledgments}

\end{article}


\begin{thebibliography}{47}
\providecommand{\natexlab}[1]{#1}
\expandafter\ifx\csname urlstyle\endcsname\relax
  \providecommand{\doi}[1]{doi:\discretionary{}{}{}#1}\else
  \providecommand{\doi}{doi:\discretionary{}{}{}\begingroup
  \urlstyle{rm}\Url}\fi

\bibitem[{\textit{Albert and Barabasi}(2002)}]{albert02}
Albert, R., and A.-L. Barabasi (2002), Statistical mechanics of complex
  networks, \textit{Review of Modern Physics}, \textit{74}, 47.

\bibitem[{\textit{Baiesi and Paczuski}(2005)}]{baiesi05}
Baiesi, M., and M.~Paczuski (2005), Complex networks of earthquakes and
  aftershocks, \textit{Nonlin. Proc. Geophys.}, \textit{12}, 1.

\bibitem[{\textit{Bakun et~al.}(2005)}]{bakun05}
Bakun, W.~H., et~al. (2005), Implications for prediction and hazard assessment
  from the 2004 parkfield earthquake, \textit{Nature (London)}, \textit{437},
  969.

\bibitem[{\textit{B{\aa}th}(1965)}]{bath65}
B{\aa}th, M. (1965), Lateral inhomogeneities in the upper mantle,
  \textit{Tectonophysics}, \textit{2}, 483.

\bibitem[{\textit{Corral}(2004)}]{corral04}
Corral, A. (2004), Long-term clustering, scaling, and universality in the
  temporal occurrence of earthquakes, \textit{Physical Review Letters},
  \textit{92}, 108,501.

\bibitem[{\textit{Corral}(2006)}]{corral06}
Corral, A. (2006), Universal earthquake-occurrence jumps, correlations with
  time, and anomalous diffusion, \textit{Physical Review Letters}, \textit{97},
  178,501.

\bibitem[{\textit{Davidsen and Goltz}(2004)}]{davidsen04}
Davidsen, J., and C.~Goltz (2004), Are seismic waiting time distributions
  universal?, \textit{Geophysical Research Letters}, \textit{31}, L21,612.

\bibitem[{\textit{Davidsen and Paczuski}(2005)}]{davidsen05m}
Davidsen, J., and M.~Paczuski (2005), Analysis of the spatial distribution
  between successive earthquakes, \textit{Physical Review Letters},
  \textit{94}, 048,501.

\bibitem[{\textit{Davidsen et~al.}(2006)\textit{Davidsen, Grassberger, and
  Paczuski}}]{davidsen05pm}
Davidsen, J., P.~Grassberger, and M.~Paczuski (2006), Earthquake recurrence as
  a record breaking process, \textit{Geophysical Research Letters},
  \textit{33}, L11,304.

\bibitem[{\textit{Davidsen et~al.}(2008)\textit{Davidsen, Grassberger, and
  Paczuski}}]{davidsen06pm}
Davidsen, J., P.~Grassberger, and M.~Paczuski (2008), Networks of recurrent
  events, a theory of records, and application to finding causal signatures in
  seismicity, \textit{Physical Review E}, \textit{77}, 066,104.

\bibitem[{\textit{Felzer and Brodsky}(2006)}]{felzer06}
Felzer, K.~R., and E.~E. Brodsky (2006), Decay of aftershock density with
  distance indicates triggering by dynamic stress, \textit{Nature (London)},
  \textit{441}, 735.

\bibitem[{\textit{Fischer and Hor\'alek}(2005)}]{fischer05}
Fischer, T., and J.~Hor\'alek (2005), Slip-generated patterns of swarm
  microearthquakes from {W}est {B}ohemia/{V}ogtland (central {E}urope):
  evidence of their triggering mechanism?, \textit{Journal of Geophysical
  Research}, \textit{110}, B05S21.

\bibitem[{\textit{Forsyth et~al.}(2009)\textit{Forsyth, Lay, Aster, and
  Romanowicz}}]{forsyth09}
Forsyth, D.~W., T.~Lay, R.~C. Aster, and B.~Romanowicz (2009), Grand challenges
  for seimology, \textit{EOS, Transactions, AGU}, \textit{90}, 361.

\bibitem[{\textit{Gomberg and Felzer}(2008)}]{gomberg08}
Gomberg, J., and K.~Felzer (2008), A model of earthquake triggering
  probabilities and application to dynamic deformations constrained by ground
  motion observations, \textit{Journal of Geophysical Research}, \textit{113},
  B10,317.

\bibitem[{\textit{Gutenberg and Richter}(1949)}]{gutenberg}
Gutenberg, B., and C.~Richter (1949), \textit{Seismicity of the Earth},
  Princeton University Press, Princeton, NJ.

\bibitem[{\textit{Hainzl and Marsan}(2008)}]{hainzl08}
Hainzl, S., and D.~Marsan (2008), Dependence of the {O}mori-{U}tsu law
  parameters on main shock magnitude: {O}bservation and modeling,
  \textit{Journal of Geophysical Research}, \textit{113}, B10,309.

\bibitem[{\textit{Hainzl et~al.}(2006)\textit{Hainzl, Z\"oller, and
  Main}}]{hainzl06}
Hainzl, S., G.~Z\"oller, and I.~Main (2006), Introduction to special issue:
  Dynamics of seismicity patterns and earthquake triggering,
  \textit{Tectonophysics}, \textit{424}, 135.

\bibitem[{\textit{Helmstetter and
  Sornette}(2002{\natexlab{a}})}]{helmstetter02}
Helmstetter, A., and D.~Sornette (2002{\natexlab{a}}), Diffusion of epicenters
  of earthquake aftershocks, omori's law, and generalized continuous-time
  random walk models, \textit{Physical Review E}, \textit{66}, 061,104.

\bibitem[{\textit{Helmstetter and
  Sornette}(2002{\natexlab{b}})}]{helmstetter_diffusion_2002}
Helmstetter, A., and D.~Sornette (2002{\natexlab{b}}), Diffusion of epicenters
  of earthquake aftershocks, omori's law, and generalized continuous-time
  random walk models, \textit{Phys. Rev. E}, \textit{66}, 061,104.

\bibitem[{\textit{Helmstetter et~al.}(2005)\textit{Helmstetter, Kagan, and
  Jackson}}]{helmstetter05}
Helmstetter, A., Y.~Y. Kagan, and D.~D. Jackson (2005), Importance of small
  earthquakes for stress transfers and earthquake triggering, \textit{Journal
  of Geophysical Research}, \textit{110}, B05S08.

\bibitem[{\textit{Helmstetter et~al.}(2006)\textit{Helmstetter, Kagan, and
  Jackson}}]{helmstetter06}
Helmstetter, A., Y.~Y. Kagan, and D.~D. Jackson (2006), Comparison of
  short-term and time-independent earthquake forecast models for southern
  california, \textit{Bulletin of the Seismological Society of America},
  \textit{96}, 90.

\bibitem[{\textit{Kagan}(2002)}]{kagan02}
Kagan, Y.~Y. (2002), Aftershock zone scaling, \textit{Bulletin of the
  Seismological Society of America}, \textit{92}, 641.

\bibitem[{\textit{Kagan}(2004)}]{kagan04}
Kagan, Y.~Y. (2004), \textit{Bulletin of the Seismological Society of America},
  \textit{94}, 1207.

\bibitem[{\textit{Kagan and Houston}(2005)}]{kagan05}
Kagan, Y.~Y., and H.~Houston (2005), Relation between mainshock rupture process
  and {O}mori's law for aftershock moment release rate, \textit{Geophysical
  Journal International}, \textit{163}, 1039.

\bibitem[{\textit{{Kagan} and {Knopoff}}(1987)}]{kagan_statistical_1987}
{Kagan}, Y.~Y., and L.~{Knopoff} (1987), Statistical {Short-Term} earthquake
  prediction, \textit{Science}, \textit{236}(4808), 1563--1567,
  \doi{10.1126/science.236.4808.1563}.

\bibitem[{\textit{Kanamori and Brodsky}(2004)}]{kanamori04}
Kanamori, H., and E.~E. Brodsky (2004), The physics of earthquakes,
  \textit{Reports on Progress in Physics}, \textit{67}, 1429.

\bibitem[{\textit{Lennartz et~al.}(2008{\natexlab{a}})\textit{Lennartz, Bunde,
  and Turcotte}}]{lennartz08}
Lennartz, S., A.~Bunde, and D.~L. Turcotte (2008{\natexlab{a}}), Missing data
  in aftershock sequences: explaining the dviations from scaling laws,
  \textit{Physical Review E}, \textit{78}, 041,115.

\bibitem[{\textit{Lennartz et~al.}(2008{\natexlab{b}})\textit{Lennartz, Bunde,
  and Turcotte}}]{lennartz_missing_2008}
Lennartz, S., A.~Bunde, and D.~L. Turcotte (2008{\natexlab{b}}), Missing data
  in aftershock sequences: Explaining the deviations from scaling laws,
  \textit{Physical Review E}, \textit{78}(4), 041,115--8,
  \doi{10.1103/PhysRevE.78.041115}.

\bibitem[{\textit{Main}(2006)}]{main06}
Main, I. (2006), A hand on the aftershock trigger, \textit{Nature (London)},
  \textit{441}, 704.

\bibitem[{\textit{Marsan and Lenglin\'e}(2008)}]{marsan08}
Marsan, D., and O.~Lenglin\'e (2008), Extending earthquake' reach through
  cascading, \textit{Science}, \textit{319}, 1076.

\bibitem[{\textit{Narteau et~al.}(2009)\textit{Narteau, Byrdina, Shebalin, and
  Schorlemmer}}]{narteau09}
Narteau, C., S.~Byrdina, P.~Shebalin, and D.~Schorlemmer (2009), Common
  dependence on stress for the two fundamental laws of statistical seismology,
  \textit{Nature (London)}, \textit{462}, 642.

\bibitem[{\textit{Newman}(2003)}]{newman03}
Newman, M. E.~J. (2003), The structure and function of complex networks,
  \textit{SIAM Review}, \textit{45}, 167.

\bibitem[{\textit{Ogata}(1988)}]{ogata_statistical_1988}
Ogata, Y. (1988), Statistical models for earthquake occurrences and residual
  analysis for point processes, \textit{Journal of the American Statistical
  Association}, \textit{83}(401), 9--27.

\bibitem[{\textit{Ogata and Zhuang}(2006)}]{ogata06}
Ogata, Y., and J.~Zhuang (2006), Space-time {ETAS} models and an improved
  extension, \textit{Tectonophysics}, \textit{413}, 13.

\bibitem[{\textit{Peixoto and Davidsen}(2008)}]{peixoto07}
Peixoto, T.~P., and J.~Davidsen (2008), Network of recurrent events in the
  {O}lami-{F}eder-{C}hristensen model, \textit{Physical Review E}, \textit{77},
  066,107.

\bibitem[{\textit{Peng et~al.}(2006)\textit{Peng, Vidale, and
  Houston}}]{peng06}
Peng, Z., J.~E. Vidale, and H.~Houston (2006), Anomalous early aftershock decay
  rate of the 2004 {Mw6.0} {P}arkfield, {C}alifornia, earthquake,
  \textit{Geophysical Research Letters}, \textit{33}, L17,307.

\bibitem[{\textit{Rubin}(2002)}]{rubin02}
Rubin, A.~M. (2002), Aftershocks of microearthquakes as probes of the mechanics
  of rupture, \textit{Journal of Geophysical Research}, \textit{107}, 2142.

\bibitem[{\textit{Rundle et~al.}(2003)\textit{Rundle, Turcotte, Shcherbakov,
  Klein, and Sammis}}]{rundle03}
Rundle, J.~B., D.~L. Turcotte, R.~Shcherbakov, W.~Klein, and C.~Sammis (2003),
  Statistical physics approach to understanding the multiscale dynamics of
  earthquake fault systems, \textit{Review of Geophysics}, \textit{41}, 1019.

\bibitem[{\textit{Shcherbakov et~al.}(2004)\textit{Shcherbakov, Turcotte, and
  Rundle}}]{shcherbakov04}
Shcherbakov, R., D.~L. Turcotte, and J.~B. Rundle (2004), A generalized omori's
  law for earthquake aftershock decay, \textit{Geophysical Research Letters},
  \textit{31}, L11,613.

\bibitem[{\textit{Shcherbakov et~al.}(2006)\textit{Shcherbakov, Turcotte, and
  Rundle}}]{shcherbakov06}
Shcherbakov, R., D.~L. Turcotte, and J.~B. Rundle (2006), Scaling properties of
  the parkfield aftershock sequence, \textit{Bulletin of the Seismological
  Society of America}, \textit{96}, 376.

\bibitem[{\textit{Sornette and Werner}(2005{\natexlab{a}})}]{sornette05w}
Sornette, D., and M.~J. Werner (2005{\natexlab{a}}), Constraints on the size of
  the smallest triggering earthquake from the epidemic-type aftershock sequence
  model, b{\aa}th's law, and observed aftershock sequences, \textit{Journal of
  Geophysical Research}, \textit{110}, B08,304.

\bibitem[{\textit{Sornette and Werner}(2005{\natexlab{b}})}]{sornette05w_a}
Sornette, D., and M.~J. Werner (2005{\natexlab{b}}), Apparent clustering and
  apparent background earthquakes biased by undetected seismicity,
  \textit{Journal of Geophysical Research}, \textit{110}, B09,303.

\bibitem[{\textit{Turcotte et~al.}(2007)\textit{Turcotte, Holliday, and
  Rundle}}]{turcotte07}
Turcotte, D.~L., J.~R. Holliday, and J.~B. Rundle (2007), {BASS}, an
  alternative to {ETAS}, \textit{Geophysical Research Letters}, \textit{34},
  L12,303.

\bibitem[{\textit{Utsu et~al.}(1995)\textit{Utsu, Ogata, and
  Matsu'ura}}]{utsu95}
Utsu, T., Y.~Ogata, and R.~S. Matsu'ura (1995), The centenary of the omori
  formula for a decay law of aftershock activity, \textit{J. Phys. Earth},
  \textit{43}, 1.

\bibitem[{\textit{Vere-Jones}(2005)}]{vere-jones05}
Vere-Jones, D. (2005), A class of self-similar random measure, \textit{Advances
  in {A}pplied {P}robability}, \textit{37}, 908.

\bibitem[{\textit{Wells and Coppersmith}(1994)}]{wells94}
Wells, D.~L., and K.~J. Coppersmith (1994), New empirical relationships between
  magnitude, rupture length, rupture width, rupture area, and surface
  displacement, \textit{Bulletin of the Seismological Society of America},
  \textit{84}, 974.

\bibitem[{\textit{Zaliapin et~al.}(2008)\textit{Zaliapin, Gabrielov,
  Keilis-Borok, and Wong}}]{zaliapin08}
Zaliapin, I., A.~Gabrielov, V.~Keilis-Borok, and H.~Wong (2008), Clustering
  analysis of seismicity and aftershock identification, \textit{Physical Review
  Letters}, \textit{101}, 018,501.

\end{thebibliography}
\end{document}